\documentclass[aps,prb,twocolumn,showpacs,amsmath,amssymb,floatfix, superscriptaddress]{revtex4-1}

\usepackage[utf8]{inputenc}

\usepackage{graphicx}
\usepackage{bm}
\usepackage{color}
\usepackage{epstopdf}
\usepackage{units}
\usepackage{natbib}
\usepackage{ulem}
\begin{document}

\author{Lo\" ic Henriet}
\affiliation{Center for Theoretical Physics, Ecole Polytechnique, CNRS, 91128 Palaiseau, France}
\author{Andrew N. Jordan}
\affiliation{Department of Physics and Astronomy, University of Rochester, Rochester, New York 14627, USA}
\affiliation{Institute for Quantum Studies, Chapman University, 1 University Drive, Orange, CA 92866, USA}
\author{Karyn Le Hur}
\affiliation{Center for Theoretical Physics, Ecole Polytechnique, CNRS, 91128 Palaiseau, France}

\title{Electrical current from quantum vacuum fluctuations in nano-engines}

\date{\today}

% Some custom definitions to save writing
% Ensemble shorthand

\newcommand{\mean}[1]{\langle #1 \rangle}           % Mean
\newcommand{\cmean}[2]{\,_{#1}\langle #2 \rangle}   % Conditioned Mean

% Dirac notation shorthand
\newcommand{\ket}[1]{|#1\rangle}                    % Ket
\newcommand{\bra}[1]{\langle #1|}                   % Bra
\newcommand{\ipr}[2]{\langle #1 | #2 \rangle}       % Inner Product
\newcommand{\opr}[2]{\ket{#1}\bra{#2}}              % Dyadic Product
\newcommand{\pprj}[1]{\opr{#1}{#1}}                 % Pure Projector

% Operator shorthand
\newcommand{\Tr}[1]{\mbox{Tr}\left[#1\right]}       % Trace
\newcommand{\comm}[2]{\left[#1,\,#2\right]}         % Commutator
\newcommand{\acomm}[2]{\left\{#1,\,#2\right\}}      % Anti-commutator
\def\R{\mbox{Re}}                                   % Real Part
\newcommand{\op}[1]{\hat{#1}}                       % Operator
\def\prj{\op{\Pi}}                                  % Projector

% Other Objects
\newcommand{\oper}[1]{\mathcal{#1}}                 % Property
\newcommand{\prop}[1]{\textit{#1}}                  % Operation
\def\gbar{\bar{\gamma}}
\def\ebar{\bar{\eta}}
\def\be{\begin{equation}}
\def\ee{\end{equation}}
\def\la{\langle}
\def\ra{\rangle}
\begin{abstract}
We theoretically investigate a quantum dot coupled to fermionic (electronic) leads and show how zero-point quantum fluctuations stemming from bosonic environments permit the rectification of the current. The bosonic baths are either external impedances modeled as tunable transmission lines or LC resonators (single-mode cavities). Voltage fluctuations stemming from the external impedances at zero temperature are described through harmonic oscillators (photon-like excitations) then producing the quantum vacuum fluctuations. The differing sizes of the zero-point fluctuations of the quantum vacuum break the spatial symmetry of the system if the quantum dot is coupled to two reservoirs or two junctions with different bosonic environments. We consider current rectification and power production when the system is operated as a heat engine in both non-resonant and resonant sequential tunneling cases.
\end{abstract}
\maketitle

\section{Introduction}
The dream of using fluctuations from the quantum vacuum as a limitless power source has inspired many, from science fiction writers, to scientists. While it is certainly possible to extract energy from the vacuum (a pair of Casimir plates moving toward each other can do external work \cite{Forward}), the difficult part is to find a way to do it cyclically or continuously without putting in more energy than you get out. Indeed, if we consider the quantum vacuum as the many body ground state, then it is clear that if the Hamiltonian itself is not changed, no energy can come from it. Quantum vacuum fluctuations arise simply because the operators in question (position, momentum, particle number, etc.) do not commute with the many body Hamiltonian, and are no longer good quantum numbers.  

We consider the possibility of rectification of electrical current from quantum vacuum fluctuations for the specific case of electrical transport through a quantum dot, and its use as a nano-heat engine. The subject of quantum thermodynamics has been rapidly growing in interest \cite{Jukka}. This has been driven both by fundamental questions as well as possible applications for energy harvesting and cooling.  There has been much recent activity on thermoelectrics and quantum dots \cite{Rafa,Sothmann2012,Jordan2013,Bjoern,Casati,Dutt_LeHur,Lim} and also pioneering experimental works \cite{experi1,experi2,experi3,experi4}. 

\begin{figure}
\includegraphics[width=3.9cm]{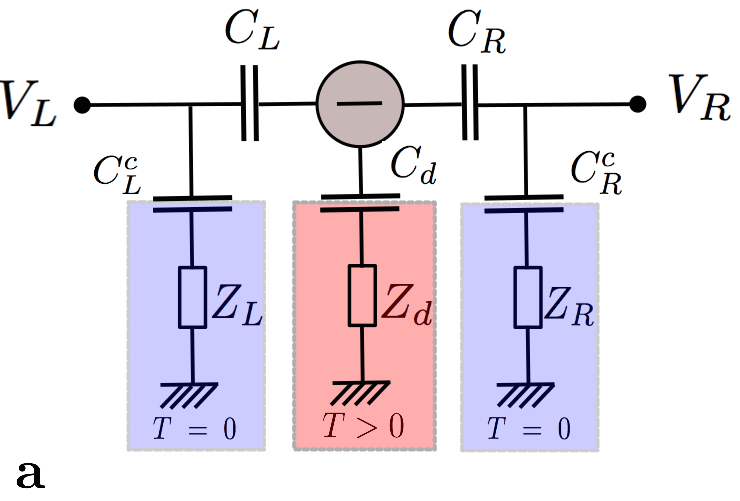}
\qquad
\includegraphics[width=3.9cm]{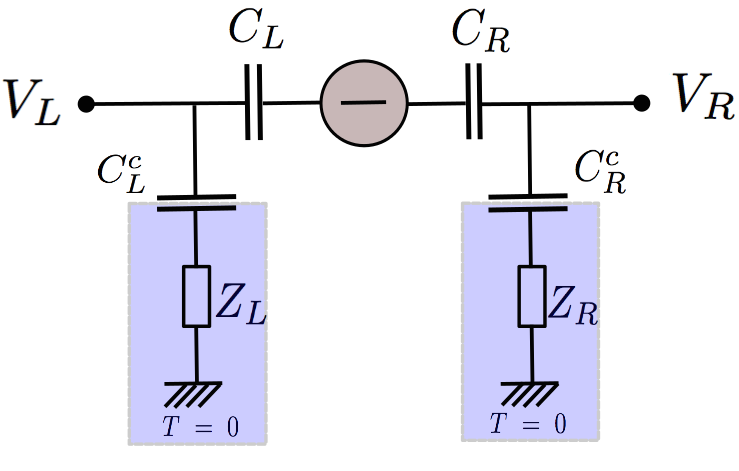}
~\\
\begin{centering}
\parbox{3.9cm}{
\includegraphics[width=3.9cm]{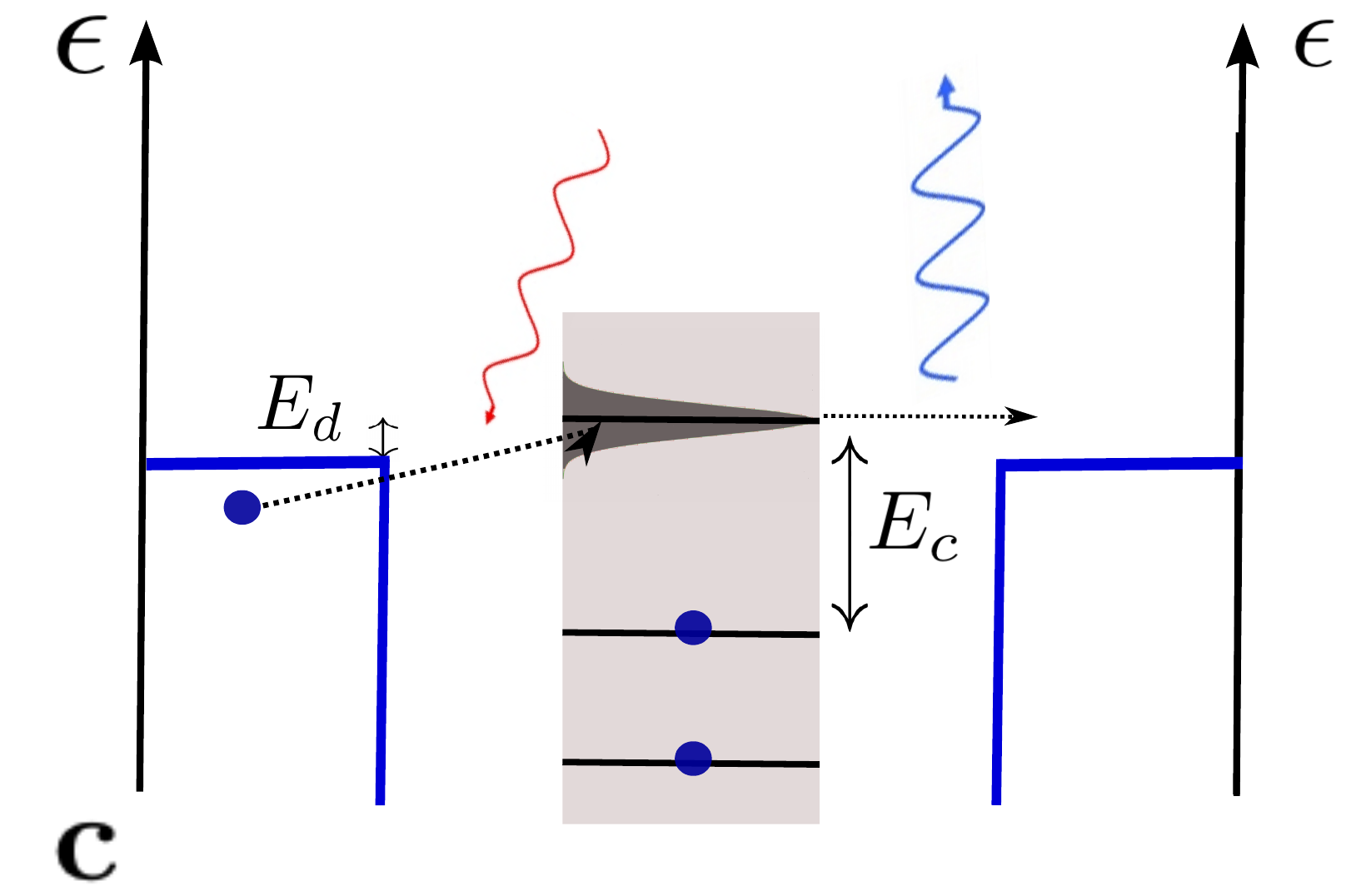}
\label{fig:2figsA}}
\qquad
\begin{minipage}{3.9cm}
\includegraphics[width=3.9cm]{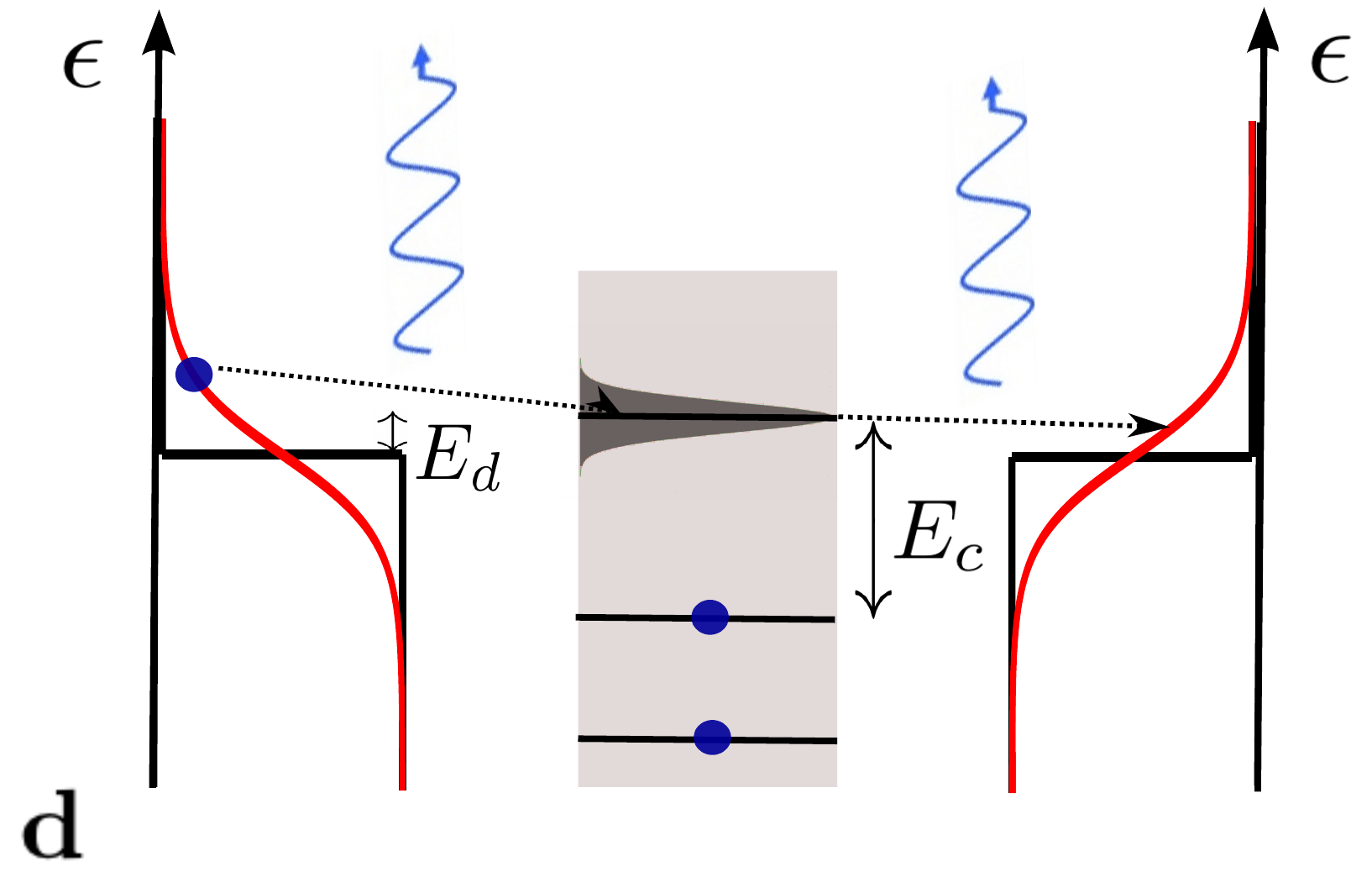}
\label{fig:2figsB}
\end{minipage}
\end{centering}
\caption{(color online). Quantum vacuum nano-engines with a quantum dot in the Coulomb blockade regime. The panels (a) and (b) represent the circuits of the two distinct setups. The panels (c) and (d) are pictorial representations of the tunneling processes respectively for the setup (a) and (b): the lateral parts correspond to the energy diagram of the electrons in the corresponding lead (at zero temperature for (c) and at non-zero temperature for (d)) and the central part represents the energy levels of the quantum dot. The straight dotted arrows represent  a tunneling event from left to right, associated to curvy arrows which correspond to photon emission/absorption. (a) and (c) Photons from a hot impedance $Z_d$ excite zero-temperature electrons tunneling onto the quantum dot. These electrons emit photons in the cold impedances to tunnel off of the quantum dot. (b) and (d) Electrons at finite temperature tunnel through the quantum dot and emit photons in the cold impedances to tunnel on and off of the quantum dot. In both cases, the asymmetry in the quantum vacuum fluctuations from the zero temperature latteral impedances leads to rectified current.
}
\label{figsetup}
\end{figure}

In this work, we consider the systems displayed in Fig.~1(a,b) and show how the zero-point quantum fluctuations can break the symmetry between left and right then producing a direct current without the application of a bias voltage across the system.
The left and right leads contain electrons, tunnel coupled to one another via a quantum dot. The dot is described with coupling capacitances $C_L, C_R$. Additionally, this system is part of an electrical circuit with finite impedances $Z_L$ and $Z_R$ connected to the system via the capacitances $C_L^c$ and $C_R^c$ respectively (and an additional impedance $Z_d$ coupled to the dot through the capcitance $C_d$ in the case of the setup of Fig.~1a). The impedances can be modeled by a chain of LC oscillators, which when quantized, behave as a system of quantum harmonic oscillators, whose frequency $\omega_i$ depends on the impedance and inductance of the chain. The vacuum energy of the chain is $E_0 = (1/2) \sum_i \hbar \omega_i$, and leads to fluctuations of the charge (or voltage) on the plate of the coupling capacitor to the quantum dot \cite{Milonni}.  We will address resonant and non-resonant sequential tunneling processes, associated to the emission of photons in the zero-temperature transmission lines, and we assume spin-polarized electrons for simplicity. For the regimes we consider, the spin just multiplies by two the number of conducting channels. The distinct regime of large impedances (dynamical Coulomb blockade) was addressed in Ref.~\onlinecite{RO}. We will also introduce a different setup, where the leads are now coupled to frequency selective cavities (or tunable LC reonators which can be built with current technology \cite{LC_resonators}). These two classes of bosonic environments coupling to mesoscopic circuits have attracted some attention both experimentally \cite{SPEC,Pierre, Gleb,Hofheinz,parla,Dousse,cQED,cavity1,cavity2,cavity3,cavity4} and theoretically \cite{Nano_engine_hybrid_cavity,Souquet,Cottet,hartle_kulkarni,Schiro,Karyn_1,Safi_Saleur,Chung_LeHur,Liu_Zheng}. Exotic many-body physics has also been addressed \cite{many_body_1,many_body_2,many_body_3,many_body_4,many_body_5,many_body_6,many_body_7}.

We now briefly discuss the physics of our setup. Naturally, no current is rectified when the system is put in state of overall zero temperature: it is necessary to give some energy to the system for a current to flow. We consider two ways to do this:  first, we consider the case when the conduction electrons are at zero temperature so that the charging energy of the quantum dot forbids electrons to tunnel onto it. However, if the quantum dot is capacitively coupled to an electrical environment via capacitance $C_d$ with impedance $Z_d$, and temperature $T$, the electrons have a new way of gaining energy: photons from the hot impedance can be absorbed, allowing the electrons to tunnel onto the dot. This corresponds to the setup displayed in the Fig.~1a. Second, we consider a model where the electrons in the two leads are in thermal equilibrium, both at the same small temperature $T$. This corresponds to the setup displayed in the Fig.~1b. In both cases, the finite charging energy of the quantum dot requires an energy $E$ from the incoming electron to be occupied. In the first case, this is supplied by absorbing a photon from the hot circuit; in the second it is from the fact that some of the electrons are thermally populated into that energy. Giving some energy to the system is however not sufficient to induce a current. We also need to break the spatial symmetry of the system in order to favor one direction of tunneling events with respect to the other. This can be done by coupling the leads to zero temperature electrical circuits with different impedances $Z_L$ and $Z_R$ (or the same impedance but with different coupling capacitances $C_L$ and $C_R$). To leave the dot, the electron must reenter the Fermi sea of the quasi-particles surrounding it. This process is sensitive to the vacuum fluctuations of the electrical circuit. If the impedances coupled to the two leads are different, the fluctuations of the vacuum will break Left/Right symmetry, giving a preferred direction for the particle to exit the system into - thus giving current rectification. \\

The paper is organized as follows. In Sec. II, we introduce the quantum many-body physics corresponding to Fig.~1, and discuss the physics in terms of $P(E)$ theory\cite{IN}. In Sec. III, we derive the quantum vacuum-induced rectified current for the different geometries considered, and show its dependence on the physical parameters of the system. In Sec. IV, we add a load voltage to the system, and consider its operation as a heat engine, finding the power produced and thermodynamic efficiency versus temperature and asymmetry. We give our conclusions in Sec. V. In Appendix A, we quantize the voltage fluctuations through a transmission line description.

\section{Analysis}
We let $c_k$ be a fermionic annihilation operator on the left lead of momentum $k$, similarly, $d$ and $c_q$ describe electrons on the dot, and the right lead. 

The system Hamiltonian is, $H = H_L + H_R + H_d+ H_{ph} + H_C$, where $H_L = \sum_k (\epsilon_k + e \delta V_L) c_k^\dagger c_k$, $H_R = \sum_q (\epsilon_q + e \delta V_R) c_q^\dagger c_q$ are the Hamiltonians for the left and right leads with electron energies $\epsilon_{k}$ and $\epsilon_{q}$. $H_{ph}$ is the Hamiltonian of the environment in isolation and $\delta V_{L,R}$ are the voltage fluctuations on the capacitors induced by the presence of the Left/Right environments. A detailled description of these terms is provided in the appendix A. We define the zero of energy as the Fermi level of the leads. The quantum dot Hamiltonian, describing a single quantum level, is $H_d =E_d+E_c (d^\dagger d-1)$. This specific form of charging energy assumes spin-polarized electrons. $E_d$ corresponds to the energy level of the dot, as defined in Fig. 1 and $E_c>0$ is the charging energy of this level, which corresponds to the difference of energy when the dot is charged, versus uncharged \cite{charging_1,charging_2}. The external impedances are coupled to the electrical circuit through coupling capacitances, allowing then an efficient thermal isolation. We find by a circuit analysis that $E_c =e^2/2C_\Sigma - e \sum_j (C_j/C_\Sigma) \delta V_j$, where the sum is over $L, R, d$. Here $C_\Sigma = C_L + C_R + C_d$ and $\delta V_d$ corresponds to the voltage fluctuations linked to the central impedance $d$, which is only present in the case of Fig.~1(a). The only effect of these impedances is then to induce environment-assisted tunneling events.

%We find by a circuit analysis that it depends on the voltages on the nearby capacitors, $E_c =e^2/2C_\Sigma - e \sum_j (C_j/C_\Sigma) \delta V_j$, where the sum is over $L, R, d$. 
%Here $C_\Sigma = C_L + C_R + C_d$ and $\delta V_d$ corresponds to the voltage fluctuations linked to the central impedance $d$, which is only present in the case of Fig.~1(a). In the two disctinct setups, the bosons couple to the total charge of the leads (or the dot).

The tunneling terms are
$H_C  = \sum_k t_{kd} d^\dagger c_k + h.c. +   \sum_q t_{dq} c_q^\dagger d + h.c.$.
Here, $t_{kd}$ is the tunneling matrix element onto the dot from the left lead with momentum $k$, $t_{dq}$ is the tunneling matrix element from the dot to the right lead with momentum $q$. The effect of the fluctuating voltages on the tunnel junctions and coupling capacitors can be taken into account by a unitary transformation $U = \exp ( i \phi_{L} \sum_k c_k^\dagger c_k + i \phi_{R} \sum_q c_q^\dagger c_q + i \phi_{d} d^\dagger d)$.
 This transformation has the effect of eliminating the fluctuating potentials on the bare system Hamiltonians, and putting a fluctuating phase on the tunneling elements,
\begin{eqnarray}
t_{kd} \rightarrow t_{kd} \exp(i \Phi_{L}),
t_{dq} \rightarrow t_{dq} \exp(- i\Phi_{R}).
%t_{kd} \rightarrow t_{kd} e^{i \phi_{L,0} - i\phi_{d,0}},
%t_{dq} \rightarrow t_{kd} e^{i \phi_{d,0} - i\phi_{R,0}}.
\end{eqnarray}
Here, we have introduced the total phase $\Phi_{l}=\phi_{l} - \phi_{d}$, where $\phi_{l} = (e/\hbar) \int_0^t dt' \delta V_l(t')$ for  $l \in \{L, R\}$, and $\phi_{d} = (e/\hbar) \sum_j (C_j/C_{\Sigma}) \int_0^t dt' \delta V_j(t')$ where the sum is over $j\in \{L, R,d\}$. The voltage fluctuations $\delta V_j$ are linked to the environmental degrees of freedom whose dynamics is governed by $H_{env}$, as shown in the appendix. Below we show how the impedances in the different circuits of Fig.~1a and Fig.~1b contribute to the noise. 

We treat the problem within $P(E)$ theory \cite{IN}, assuming that tunneling on and off of the dot can be described with tunneling rates described within a golden rule picture in perturbation theory. This also assumes that environments relax faster than the tunneling time scale rates. In this picture, we do not take into account the backaction of the electrons on the excitations of the impedances. The tunneling rates are controlled by the probability of the electrical circuit to either absorb or emit a photon. This probability for lead $l \in \{L,R\}$ is given by
\be
P_l(E) = (1/h) \int_{-\infty}^\infty dt e^{i E t/\hbar} e^{K_l(t)},
\ee
where $K_l(t) = \la (\Phi_l(t) - \Phi_l(0)) \Phi_l(0)\ra$ (where the $\Phi_l$ variables are defined above).  \\

In the case of cold electrons and hot photons (Fig.~1(a)) we have $K_l = (\eta_{ll}^2 + \eta_{lm}^2) k_l + \eta_d^2 k_{d}$ for $(l,m)=(L,R),(R,L)$ where we have introduced the coupling constants to the environmental baths, $\eta_{ll} = (C_{m} + C_d)/C_\Sigma, \eta_{lm} = C_{m}/C_\Sigma, \eta_d = C_d/C_\Sigma$ ($m=R$ when $l=L$ and vice versa). $k_l(t)$ and $k_d(t)$ are the lead and dot correlation functions taken in isolation $k_j(t) = \la (\phi_j(t) - \phi_j(0)) \phi_j(0)\ra$ and they read 
\begin{align}
k_l(t)&=\int_{0}^{\infty} \frac{d \omega}{\omega} 2 Re  \left[ \frac{Z_l}{ R_q}\right]  (e^{-i\omega t}-1), \notag \\
k_d(t)&=\int_{0}^{\infty} \frac{d \omega}{\omega} 2 Re  \left[ \frac{Z_d}{ R_q} \right] \Big\{\coth\left(\frac{\beta \hbar \omega}{2}\right)\left[\cos(\omega t)-1\right]\notag \\
&\phantom{{}=111111111111111111111}-i\sin(\omega t)         \Big\},
\label{k}
\end{align}
where $Z_l=\sqrt{L_l/C^c_l}$ is the impedance coupled to the lead $l$. $Z_l$ is expressed as a function of $L_l$ and $C^c_l$, which are the values of the inductances and capacitances which compose the transmission line (see appendix A). Physically, the correspondence between impedance and transmission line just fixes the ratio between inductances and capacitances, therefore we can choose the capacitances to be equal to $C_l^c$ and adjust the inductances such that the ratio $\sqrt{L_l/C_l^c}$ gives the physical resistance at $\omega=0$. $R_q = h/e^2$ is the resistance quantum.

For the sake of simplicity in this case of the setup 1(a), we have specified the case for which both left and right leads have the same correlation function $k_{L}(t)=k_{R}(t)=k_{0}(t)$ in isolation at zero temperature, and the asymmetry of the system comes from different values of $C_L$ and $C_R$. This is one particular choice to break the Left/Right symmetry, which leads to the simple expression above for $K_l$. There are other ways to break this symmetry such as having different impedances $Z_L$ and $Z_R$. The important point is that the capacitive coupling to the hot impedance now enters the $P_l(E)$ functions for the tunneling electrons.\\

In the case of hot electrons and cold photons (Fig.~1(b)) we have $\eta_d=0$ and for $C_L=C_R$ we find that $K_l=k_l$ given directly by the first line of Eq. (\ref{k}). In this case we will take distinct impedances leading to different functions $K_{L}(t)$ and $K_{R}(t)$.

 \section{Results for the different geometries}
\subsection{Cold electrons and hot impedance $Z_d$ (Fig.~1(a))} 

The effects of $k_{0}(t)$ and $k_d(t)$ are quite different and we will first consider their effects independently. As stated previsouly we consider for $k_{0}(t)$ the case where the zero frequency external impedance of the $T=0$ transmission lines $Z=R=\sqrt{L/C^c}$ is small compared to the resistance quantum, $R \ll R_q = h/e^2$, and define the large $\alpha = R_q/R \gg1$ (where $L=L_L=L_R$ and $C^c=C^c_L=C^c_R$). In this case, we have for long time, $k_{0}(t) = -(2/\alpha) \ln(\alpha E_c t/\pi \hbar) + i\pi/2 + \gamma_e$, where $\gamma_e$ is the Euler constant \cite{IN}. The isolated effect of this cold impedance would lead to a $P_{0}(E)=(1/h) \int_{-\infty}^\infty dt \exp\left(i E t/\hbar\right) \exp \left[ k_0(t)\right] $ function that vanishes for negative energies, and has a power-law divergence for small positive energies, signaling the onset of the orthogonality catastrophe. In contrast for the hot impedance, still in the small resistance limit, the correlation function becomes $k_d(t) = - (Re Z_d/R_q) k_B T t/\hbar$ in the long time limit. The isolated effect of this cold impedance would lead to a normalized Lorentzian for $P_d(E)=(1/h) \int_{-\infty}^\infty dt \exp\left(i E t/\hbar\right) \exp \left[ k_d(t)\right] $ of width $\Delta = (Re Z_d/R_q) k_B T$.\cite{IN}

The combined $P_l(E)$ functions from both effects is found by taking a convolution of the two $P_0(E)$ and $P_d(E)$ functions (convolution theorem), provided the coupling constants $\eta_l$ to the various circuits are properly accounted for. This turns out to be straightforward because they can be absorbed into effective impedances $\tilde \alpha_l=\alpha /(\eta_{ll}^2 + \eta_{lm}^2) $, charging energies ${\tilde E}_{c,l}=(\eta_{ll}^2 + \eta_{lm}^2) E_c$ and Lorentzian widths, $\tilde \Delta_l=\Delta \eta_d^2$, giving
\be
P_l(E) = \frac{\pi^{2/\tilde{\alpha}_l} e^{-2 \gamma_e/\tilde{\alpha}_l}
}{\Gamma(2/\tilde{\alpha}_l)  \sin(2\pi/\tilde{\alpha}_l ) (\tilde{\alpha}_l \tilde{E}_{c,l})^{2/\tilde{\alpha}_l}}
{\rm Im}( -i \tilde{\Delta}_l - E)^{\frac{2}{\tilde{\alpha}_l}-1}.
\ee

The Left/Right asymmetry is now solely contained in the effective parameters $\tilde \alpha_L \neq \tilde \alpha_R$. These $P_l(E)$ functions permit us to calculate the tunneling rates $T_{\pm,l}$ between the leads and the dot, from left to right (+) and right to left (-),  

\begin{eqnarray}
T_{\pm, L} (\Omega) = T_{0,L} \int d\epsilon f(\epsilon \pm \Omega \mp \mu/2) P_L(\epsilon), \label{Gamma_L}\\
T_{\pm, R} (\Omega) = T_{0,R} \int d\epsilon f(\epsilon \mp \Omega \mp \mu/2)  P_R(\epsilon)\label{Gamma_R}.
\end{eqnarray}

Here $T_{0,l} = 2\pi |t_l|^2 \nu_l V_l$ is the bare tunneling rate given in terms of the tunnel coupling $t_l$, the density of states $\nu_l$ and the volume $V_l$ of the lead, and we have included the possibility of a bias $V$ $(\mu=eV)$ on the system. $\Omega$ denotes the energy of the electron after the tunneling event. At zero temperature the Fermi distribution $f$ becomes a step function and equations (\ref{Gamma_L}) and (\ref{Gamma_R}) have a simple interpretation: the electrons in the lead have no way to gain energy without absorbing a photon from the bosonic bath if $\mu=0$. This is why the convolution with the hot impedance is important - it permits the $P(E)$ function to have some probability in the negative energy range, so a photon can be absorbed from that bath. It may then be given back to a cold bath as it tunnels left or right, as can be seen in Fig.~1(c). The asymmetry between the capacitances breaks the left-right symmetry. Equivalently we repeat that one may have taken the same value for $C_L$ and $C_R$ but with different impedances $Z_L$ and $Z_R$. In this case of Fig.~1(b) the level is dephased by a fluctuating phase $\phi_d$, so that we assume a sequential non-resonant tunneling case and the rectified current is given by \citep{RO},
\be
I =e ~\frac{T_{+,L}(E_d) T_{+,R}(E_d) - T_{-,L}(E_d) T_{-,R}(E_d)}{ T_{+,L}(E_d) + T_{+,R}(E_d) +  T_{-,L}(E_d) + T_{-,R}(E_d) }.
\label{rectified_current}
\ee

\subsection{Hot electrons and $Z_d = 0$ (Fig.~1(b))}
In this case no photons can be absorbed during the tunneling processes. However, we now consider the electrons to be thermally distributed in energy with a Fermi function so they can tunnel from lead to dot and vice versa. This symmetry can be broken by permitting the tunneling electrons to lose energy by emitting photons in either the left or right electrical circuits both tunneling on and off the dot, emitting two photons, as can be seen in Fig.~1(d). Here, the asymmetry between the cold impedances breaks the left-right symmetry leading to rectification of current.

As before we consider the case where the zero frequency external impedances of the transmission lines $Z_l=R_l$ is small compared to the resistance quantum, $R_l \ll R_q = h/e^2$, and define the large $\alpha_l = R_q/R_l \gg1$. In this case, we have again for long time, $K_{l}(t) = -(2/\alpha_l) \ln(\alpha_l E_c t/\pi \hbar) + i\pi/2 + \gamma_e$. The rectified current assuming a non-resonant sequential tunneling is then given by Eq.~(\ref{rectified_current}), with the following expression for the tunneling rates,

\begin{figure}[t!]
\center
\includegraphics[scale=0.4]{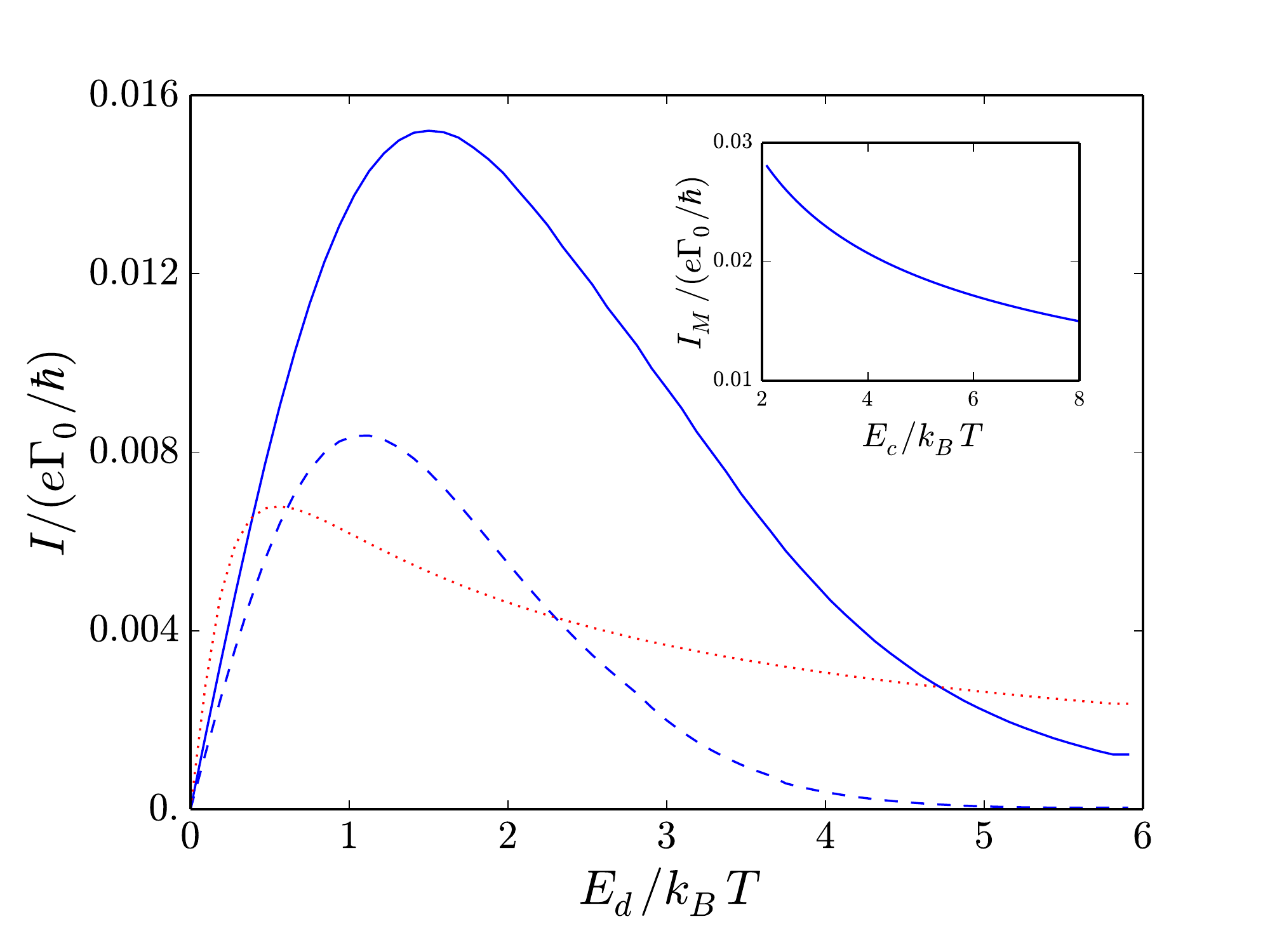}  
\caption{(color online). Rectified current with respect to $E_d/k_B T$ at zero bias. The red dotted curve corresponds to the setup displayed in Fig.~1(a), the dashed blue and the full blue curves correspond respectively to the non-resonant and the resonant case of the setup displayed in Fig.~1(b). We have $k_B T=0.125 E_c$, $\alpha_L(\tilde{\alpha}_L)=5$, $\alpha_R(\tilde{\alpha}_R)=30$, $\Gamma_0=0.125 E_c$ and $\tilde{\Delta}_l/\Gamma_0=1$. Inset: Evolution of the maximum value of the current $I_M$ at the resonance with respect to $E_c/k_B T$.}
%\label{fig_6}
\end{figure}

\begin{eqnarray}
T_{\pm,L}(\Omega)  = -T_{0,L} e^{- \frac{2\gamma_e}{\alpha_L}} \left(\frac{\pi k_B T}{\alpha_L E_c}\right)^{\frac{2}{\alpha_L}}  {\rm Li}_{\frac{2}{\alpha_L}} \left(-e^{\frac{\mp \Omega \pm \mu/2}{k_B T}}\right)\notag \\
T_{\pm,R}(\Omega)  = -T_{0,R} e^{-\frac{2\gamma_e}{\alpha_R}} \left(\frac{\pi k_B T}{\alpha_R E_c}\right)^{\frac{2}{\alpha_R}}  {\rm Li}_{{\frac{2}{\alpha_R}}} \left(-e^{\frac{\pm \Omega \pm \mu/2}{k_B T}}\right)\notag,
\end{eqnarray}
where ${\rm Li}_x(z)$ is the Polylogarithm function. $\Omega$ denotes again the energy of the electron after the tunneling event.

However, in the case of Fig.~1(b), it is probably more relevant to address resonant tunneling processes. The transmitted electron can coherently bounce back and forth between the two barriers and interfere constructively before it exits. The rectified current is now $I=(e/\hbar)\left(\Gamma^{+}-\Gamma^{-}\right)$, obtained by summing over all the possible paths of the electron, so that we have

\begin{align}
\Gamma^{+}=\int d\Omega ~  \frac{\Gamma_{+,L}(\Omega) \Gamma_{+,R}(\Omega)}{\left(\frac{\Gamma_{0,L}+\Gamma_{0,R}}{2}\right)^2+(\Omega-E_d)^2}, \\
\Gamma^{-}= \int d\Omega ~  \frac{\Gamma_{-,L}(\Omega) \Gamma_{-,R}(\Omega)}{\left(\frac{\Gamma_{0,L}+\Gamma_{0,R}}{2}\right)^2+(\Omega-E_d)^2} .
\label{Gammas_resonant}
\end{align}
We have introduced $\Gamma_{\pm,l}(\Omega)=\hbar T_{\pm,l}(\Omega)$, and we will focus on the symmetric case where $\Gamma_{0,l}=\hbar T_{0,l}=\Gamma_{0}$ in the following. This case permits Fabry-Perot type resonances between the two junctions forming the quantum dot \cite{datta}. It corresponds to a resonant sequential tunneling associated to the emission of two photons.\\

\begin{figure}[t!]
\center
\includegraphics[scale=0.4]{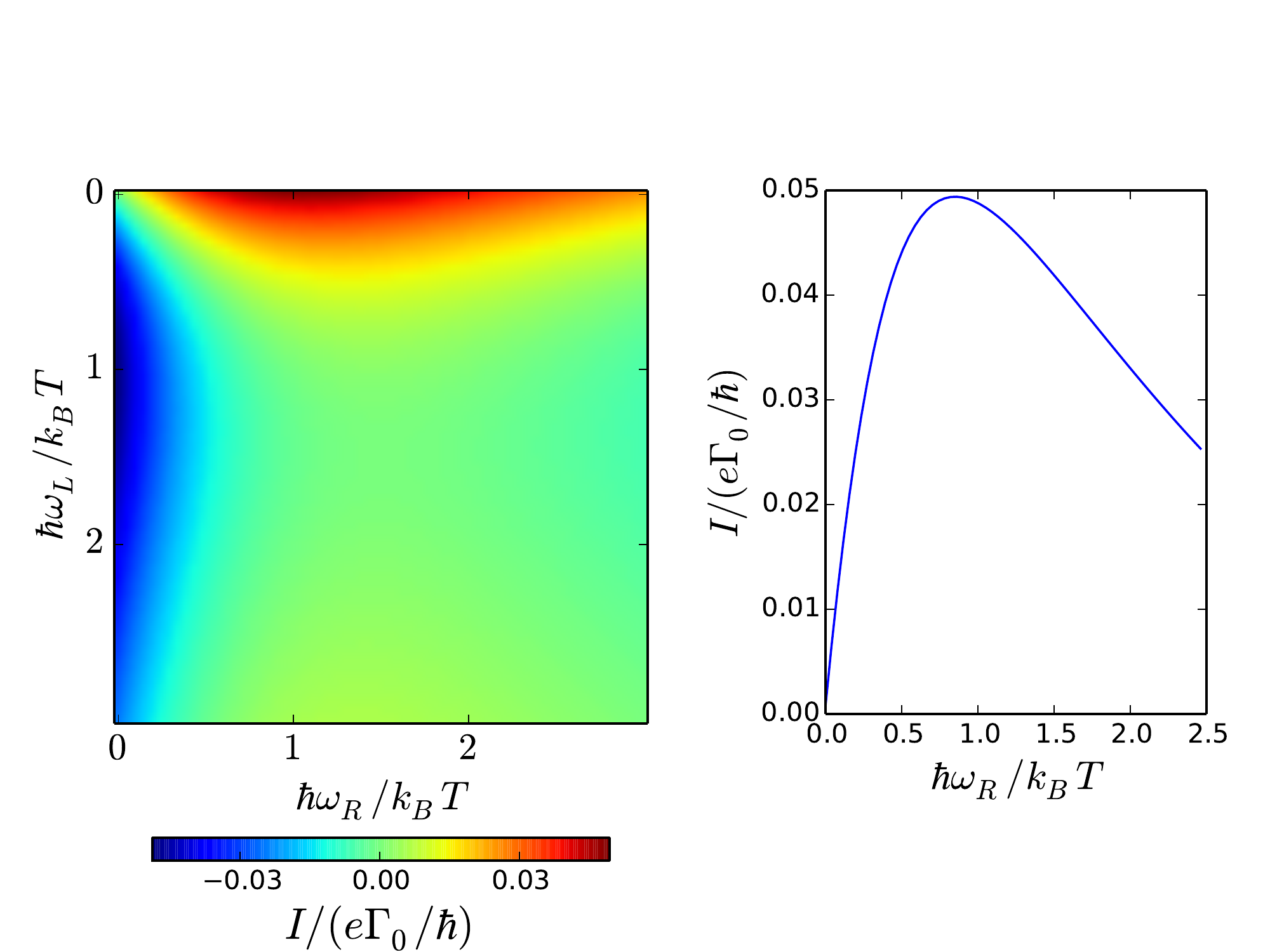}  
\caption{(color online) Left panel: Rectified current with respect to the left and right cavity frequencies $\omega_L$ and $\omega_R$. We have $k_B T/\Gamma_0=1.2$, $E_d/ \Gamma_0=2$, $Z_L/R_q=Z_R/R_q=0.5$ and $\mu=0$. Right panel: Rectified current with respect to $\omega_R$, for a setup with only one cavity coupled to one of the leads.}
%\label{fig_6}
\end{figure}

Putting these results together, we present an analysis of the evolution of the rectified current with respect to the dot energy level $E_d$ and to the charging energy $E_c$. Here we fix the temperature $T$ and $\Gamma_0$ such that $k_B T/\Gamma_0=1$. Note that $T$ is the temperature of the hot impedance for the setup displayed in Fig.~1(a), while it is the temperature of the leads for the setup displayed in Fig.~1(b). The charging energy is the highest energy scale of the problem, and we do not calculate the current at temperatures of the order or greater than the charging energy. Above this threshold, the description we made is no longer accurate as the other levels of the quantum dot must be taken into account. \\

 In Fig.~2, we plot the rectified current with respect to the dot energy for $\alpha_L(\tilde{\alpha}_L)=5$ and $\alpha_R(\tilde{\alpha}_R)=30$, at a fixed value of the charging energy $E_c/\Gamma_0=8$. We notice that this current is maximal at a dot energy of the order of the temperature. The value of this maximal current also depends on the charging energy of the dot. The inset shows then the evolution of the maximum value of the current with respect to $E_c/k_B T$ (for $E_c/k_B T>2$). We notice that the current decreases with the charging energy, as expected from the power law evolution of $P(E)$. The evolution of the current with respect to the dot energy looks different for the case of Fig.~1a and the case of Fig.~1b. These differences come from the effect of temperature, which contributes through a Bose-Einstein distribution for the case of Fig.~1a and through a Fermi distribution for the case of Fig.~1b. The latter is responsible for the sharpness of the resonance (dashed and full blue lines) while the first one leads to a rather long tail (dotted red line).

\subsection{LC circuit and cavities}

Finally we focus on the setup displayed in Fig.~1(b) where the lead $l$ is now coupled to a zero-temperature resonator of frequency $\omega_l=1/\sqrt{L_l C^c_l}$ (composed of an inductance $L_l$ and another capacitance $C^c_l$). We then have

\be
K_l(t)= \frac{Z_l}{R_q}\left(e^{-i\omega_l t}-1\right).
\ee
where we have again $Z_l= \sqrt{L_l/C^c_l}$. We find

\begin{eqnarray}
%e^{K_l(t)} =&\sum_{n=0}^{\infty} \frac{1}{n!}\left(\frac{z_l}{R_q}\right)^n \left(e^{-i\omega_l t}-1\right)^n \notag\\
e^{K_l(t)} =&\sum_{n=0}^{\infty} \sum_{k=0}^{n} \frac{1}{n!} \binom {n} {k}(-1)^{n-k} \left(\frac{Z_l}{R_q}\right)^n  e^{-i\omega_l k t}.
\end{eqnarray}
This allows us to reach the following expression for $P_l(E)$,

\begin{eqnarray}
%P_l(E)=&\sum_{n=0}^{\infty}\sum_{k=0}^{n} \frac{1}{n!} \binom {n} {k}(-1)^{n-k} \left(\frac{z_l}{R_q}\right)^n \delta(E-k\hbar\omega_l) \notag\\
%=&\sum_{k=0}^{\infty} \sum_{n=k}^{\infty} \frac{1}{n!} \binom {n} {k}(-1)^{n-k} \left(\frac{z_l}{R_q}\right)^n \delta(E-k\hbar\omega_l) \notag\\
P_l(E)=&\sum_{k=0}^{\infty} \frac{1}{k!}\left(\frac{Z_l}{R_q}\right)^k e^{-\frac{Z_l}{R_q}} \delta(E-k\hbar\omega_l).
\end{eqnarray}

\begin{figure}[t!]
\center
\includegraphics[scale=0.4]{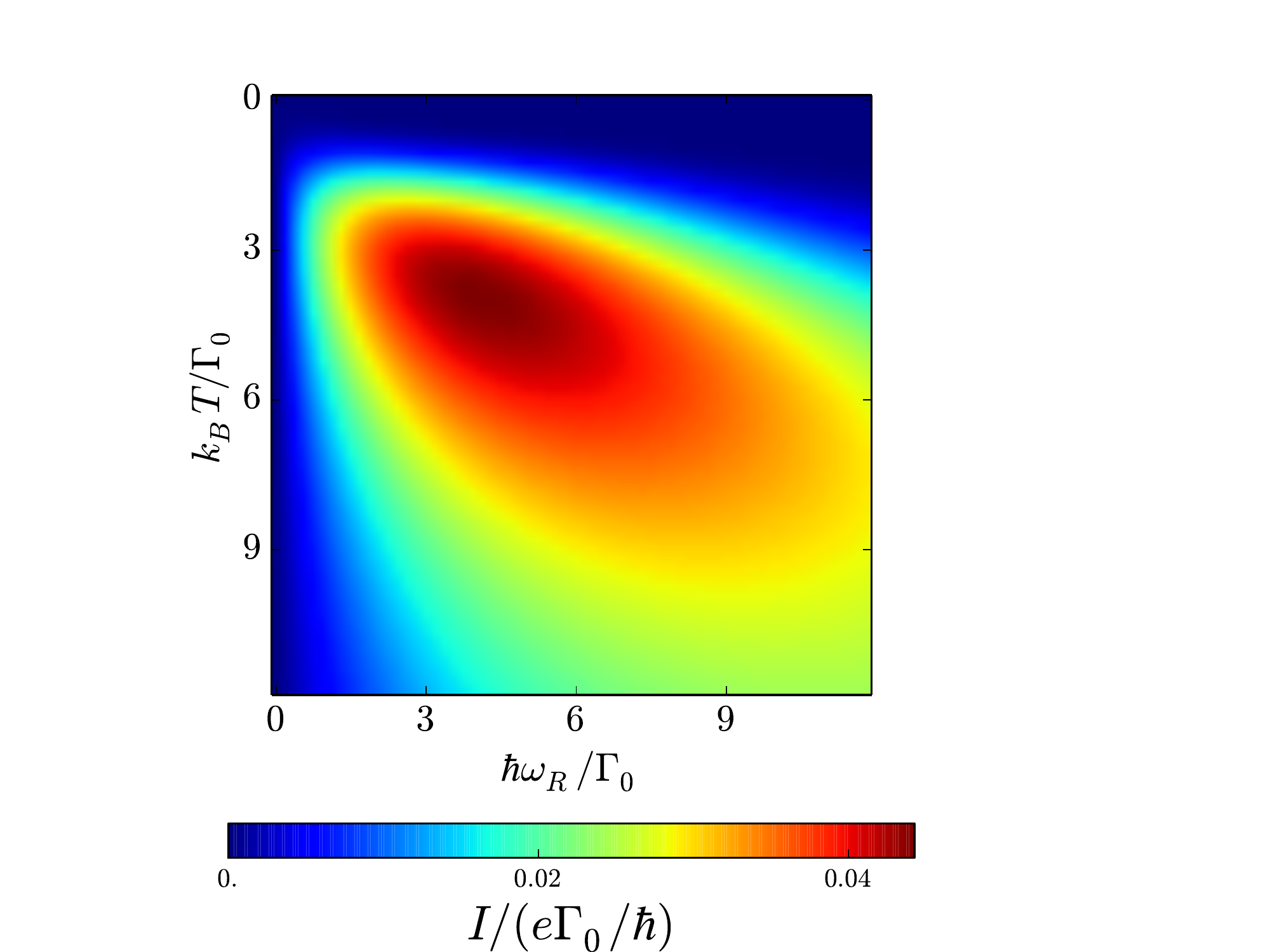}  
\caption{(color online) Rectified current with respect to the temperature, and the cavity frequency. We have $\Gamma_{0,L}=\Gamma_{0,R}=\Gamma_{0}$, $E_d/\Gamma_0=6$, $Z_R/R_q=0.5$. }
%\label{fig_6}
\end{figure}

The left and right tunneling rates given by the Eqs. (5) and (6) read

\begin{eqnarray}
T_{\pm, L} (\Omega) = T_{0,L} \sum_k  \frac{\left(\frac{Z_L}{R_q}\right)^k}{k!} e^{-\frac{Z_L}{R_q}} f(k\hbar\omega_L \pm \Omega) \\
T_{\pm, R} (\Omega) = T_{0,R} \sum_k  \frac{\left(\frac{Z_R}{R_q}\right)^k}{k!} e^{-\frac{Z_R}{R_q}} f(k\hbar\omega_R \mp \Omega).
\end{eqnarray}
These rates take into account tunneling events with the emission of several photons, and the ratio $Z_l/R_q$ determines the dominant processes occuring in the device. We first consider small values of $Z_l/R_q$ where one-photon processes are the most relevant. % and we only couple the right lead to a cavity. In this case, the current is simply given by

%\be
%I=\frac{e z_R}{\hbar R_q}\int d\Omega \frac{\Gamma_0^2}{\Gamma_0^2+(\Omega-E_d)^2} F(\Omega),
%\label{current_one_cavity_one_photon}
%\ee
%where
%\be
%F(\Omega)=\left[ f\left(\Omega\right)f\left(\hbar \omega_R-\Omega\right)-f\left(-\Omega\right)f\left(\hbar \omega_R+\Omega\right)\right].
%\ee

Here we first fix the temperature $T$ and $\Gamma_0$ such that $k_B T/\Gamma_0=1.2$. In Fig.~3 we plot the rectified current for different values of $\omega_L$ and $\omega_R$, but with constant $Z_L=Z_R$. We have then photon-assisted tunneling and current rectification without any drive on the cavity, in contrast to Refs. \onlinecite{tien_gordon,tucker,pedersen_buttiker}. The best situation corresponds to very asymmetric configurations (blue and red regions), where one of the two cavities is suppressed. In this case, the coupling to a frequency-selective cavity leads to a rectified current substantially greater than in the case of two standard resistances in a highly asymmetric configuration (see Fig.~2 and Fig.~3). \\

We now study the dependence of the rectified current with respect to the temperature and the cavity frequency in this highly asymmetric case, where only the right lead is coupled to a cavity. In this case, the current is simply given by

\be
I=\frac{e Z_R}{\hbar R_q}\int d\Omega \frac{\Gamma_0^2}{\Gamma_0^2+(\Omega-E_d)^2} F(\Omega),
\label{current_one_cavity_one_photon}
\ee
where
\be
F(\Omega)=\left[ f\left(\Omega\right)f\left(\hbar \omega_R-\Omega\right)-f\left(-\Omega\right)f\left(\hbar \omega_R+\Omega\right)\right].
\ee

 The rectified current is written in Eq. (\ref{current_one_cavity_one_photon}) as a difference of two terms, which are products of two Fermi functions. At zero temperature both terms are zero (because one of the Fermi function is zero in both terms), and no tunneling events can occur. In the limit of high temperature, the flattening of the Fermi functions erase the asymmetry of the setup introduced by $\omega_R$, which also leads to a zero rectified current. We deduce the existence of one resonance: there exists an optimal lead temperature for which current is maximal.\newline

\begin{figure}[t!]
\center
\includegraphics[scale=0.4]{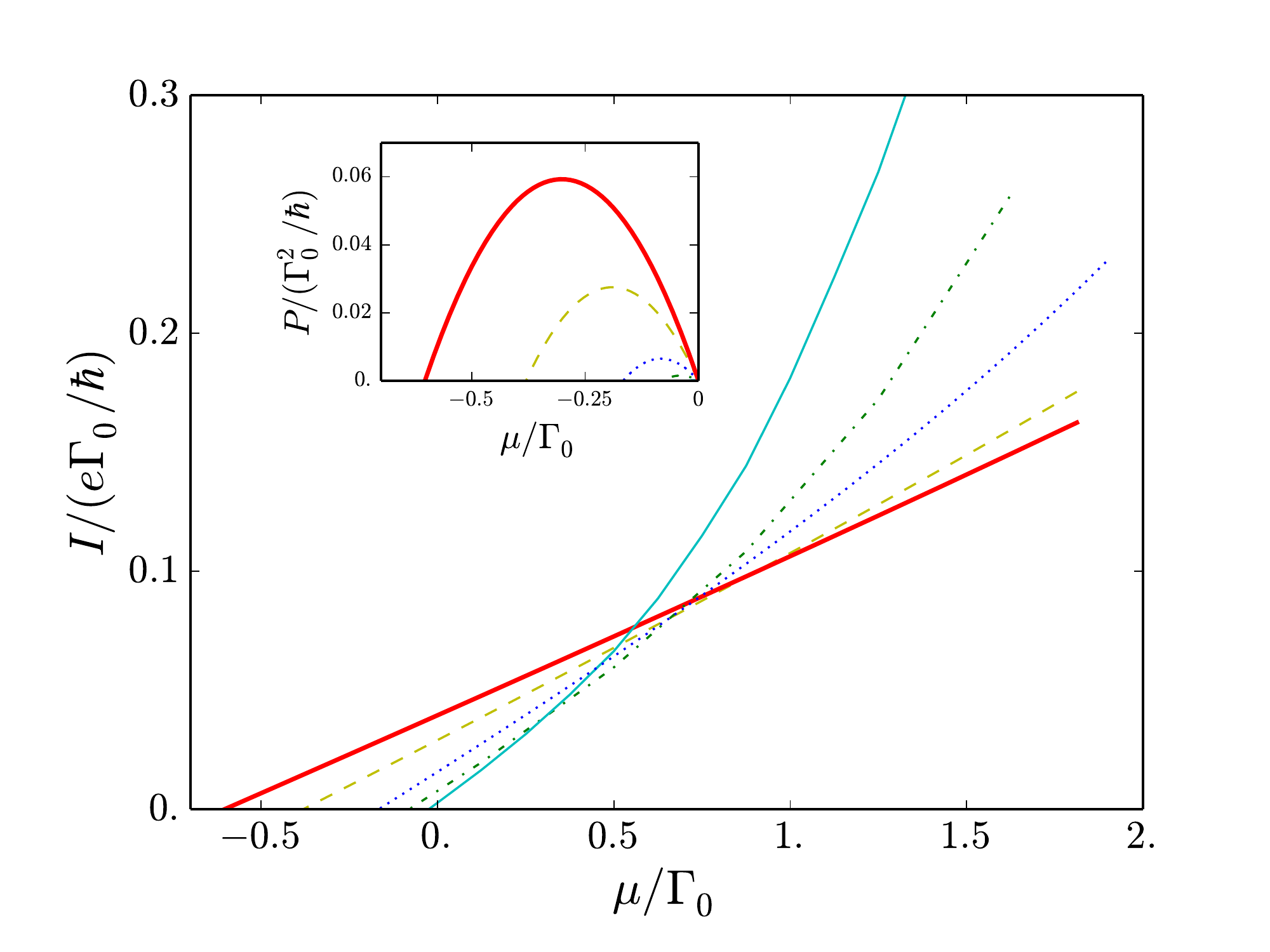}  
\caption{(color online). Rectified current with respect to an external bias at different lead temperatures, ranging linearly from the limit of zero temperature (full cyan curve) to $k_B T/\Gamma_0=3$ (full red curve). We have taken the resonant case of the setup displayed in the Fig.~1(b) (and adjusted the value of $E_d/k_B T$ at small bias). Inset: generated power with respect to the external bias. We have $\alpha_L=5$, $\alpha_R=30$ and $\Gamma_0=0.125 E_c$.}
%\label{fig_6}
\end{figure}

In Fig.~4 we plot the rectified current with respect to the temperature and to the cavity frequency. In this case of one-photon processes, we naturally find that the resonance occurs at a temperature $T$ such that $k_B T\simeq \hbar \omega_R$. For greater values of $Z_R/R_q$, this resonance is shifted to a greater value of the temperature $T$, which can be estimated by $k_B T\simeq k \hbar \omega_R$, where $k$ is the number of photons of the dominant process at this value of $Z_R/R_q$.

\section{Characteristics of heat engines}
\begin{figure}[t!]
\center
\includegraphics[scale=0.4]{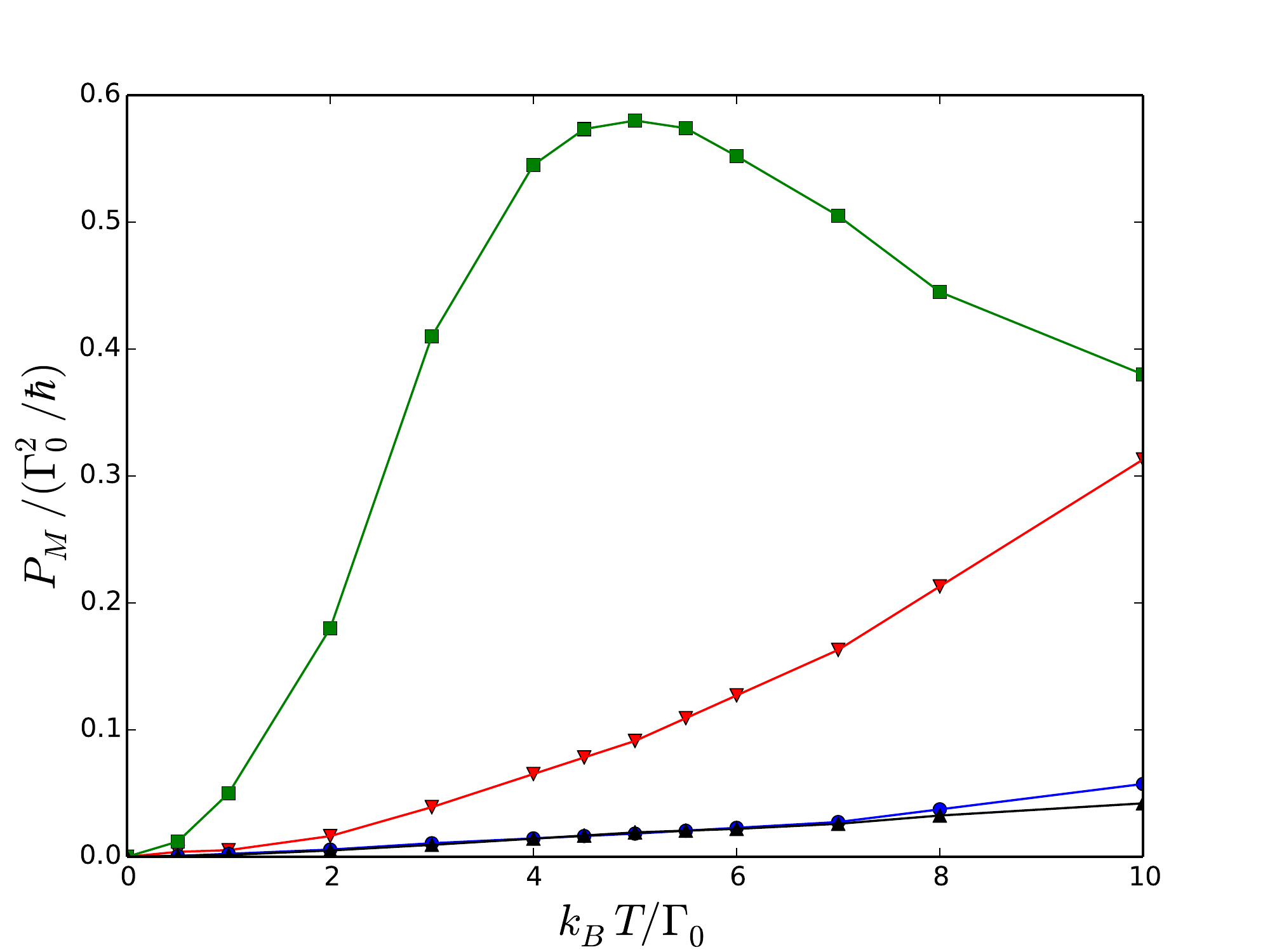}  
\caption{(color online) Maximal power $P_M$ with respect to $k_B T/\Gamma_0$. The red curve with down-pointing triangles and the blue curve with circles correspond respectively to the resonant and the non-resonant case of the setup displayed in Fig.~1(b). The black curve with up-pointing triangles correspond to the setup displayed in Fig.~1(a) with $\tilde{\Delta}_l/\Gamma_0=1$. For these setups, we have $E_c/\Gamma_0=20$, $\alpha_L (\tilde{\alpha}_L)=5$ and $\alpha_R  (\tilde{\alpha}_R)=30$, and $E_d$ is adjusted to maximize the power. The green curve with squares corresponds to the case of one cavity, and $\omega_R$ and  $E_d$ are adjusted to maximize the power.}
%\label{fig_6}
\end{figure}

\begin{figure}[h!]
\center
\includegraphics[scale=0.4]{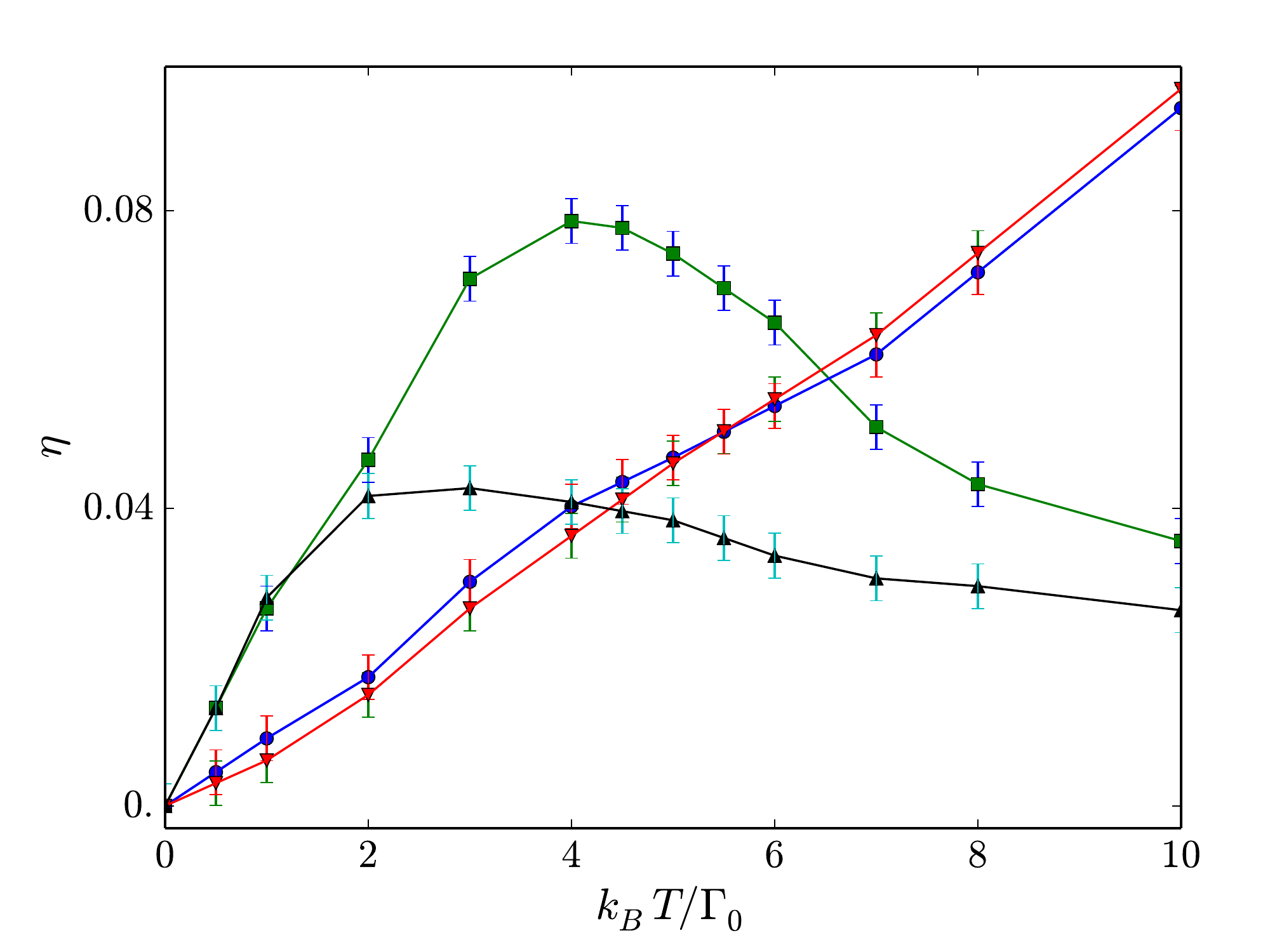}  
\caption{(color online) Efficiency of the engines with respect to $k_B T/\Gamma_0$. We use the same color and symbol legend as in Fig.~6. Parameters are unchanged.}
%\label{fig_6}
\end{figure}

The generated power is given by the product $I \times V$. When $I>0$ and $\mu<0$, the device produces power: the generated current flows against a load potential. In Fig.~5 we plot the evolution of the resonant current with respect to an external bias field at different lead temperatures, for the setup displayed in Fig.~1(b). We see that the increase of the temperature has two effects: it leads to a drop in the conductance, and to the appearance of the zero-voltage rectification current.

The heat engines can be characterized thanks to two physical quantities which are the maximum power that can be generated, whose evolution is shown in the inset of Fig.~5, and the efficiency of the engine. We plot in Fig.~6 the maximum power as a function of temperature. In the single cavity case it exhibits a peak at a given temperature, while on the other hand the power generated by the circuit coupled to the impedances is increasing with the temperature. In all cases we must consider thermal energies which are smaller than the level spacing of the dot.\\

In the case of hot leads and cold environment, the efficiency is given by $\eta=IV/(IV-J)$, where $-J$ is the heat expelled to the cold environment, which is then given by the sum of all the energy carried by the emitted photons. The heat currents associated to the tunneling events are

\begin{eqnarray}
J_{\pm, L} (\Omega) = \Gamma_{0,L} \int d\epsilon f(\epsilon \pm \Omega \mp \mu/2) P_L(\epsilon) \epsilon , \label{J_L_appendix}\\
J_{\pm, R} (\Omega) = \Gamma_{0,R} \int d\epsilon f(\epsilon \mp \Omega \mp \mu/2)  P_R(\epsilon) \epsilon \label{J_R_appendix}.
\end{eqnarray}

Defining  $J(\Omega)=J_{+, L}(\Omega)+J_{-, L}(\Omega)+J_{+, L}(\Omega)+J_{+, R}(\Omega)$, the emitted heat current is simply given by $J(E_d)$ in the case of non-resonant tunneling, while it is
\begin{align}	
J=\int d\Omega ~  J(\Omega)\frac{\left(\frac{\Gamma_{0,L}+\Gamma_{0,R}}{2}\right)}{\left(\frac{\Gamma_{0,L}+\Gamma_{0,R}}{2}\right)^2+(\Omega-E_d)^2}\notag \\
\end{align}
in the case of resonant tunneling. On the other hand, for cold leads and a hot environment, the efficiency reads $\eta=IV/J_H$, where $J_H$ is the amount of heat received from the environment. We plot in Fig.~7 the efficiency of the Nano-engines with respect to $k_B T/\Gamma_0$. We can remark that the efficiency of the setup displayed in Fig.~1(b) is similar for the non-resonant and the resonant case.

\section{Conclusion}
We have shown that the quantum vacuum fluctuations of a $T=0$ electrical circuit permit the rectification of electrical current in a mesoscopic quantum system. While external energy is required for this rectification to happen, the fact that the electrical environment absorb photons and electrons differently for different external impedances or capacitances in the system is sufficient to break the left/right spatial symmetry and rectify the current. Of course, there are other ways to break this symmetry, such as having a nonlinear electron transmission that is asymmetric (different tunneling rates is not enough). However, we stress that for our system in terms of the electron degrees of freedom, the system is completely left/right symmetric - it is the vacuum fluctuations from the bosonic degrees of freedom that break the symmetry.   

We would like to dedicate this work to Markus B\" uttiker. This work has also benefitted from discussions at the Memorial Symposium
in honor of Bernard Coqblin. We acknowledge support from the PALM Labex, Paris-Saclay,  ANR-10-LABX-0039. ANJ acknowledges support from the University of Rochester Researcher Mobility Travel Grant, and thanks Ecole Polytechnique for hospitality. We thank Olesia Dmytruk, Tal G\"{o}ren and Pascal Simon for discussions.

\appendix
\section{Voltage fluctuations and the environment}

\begin{figure}[h!]
\center
\includegraphics[scale=0.24]{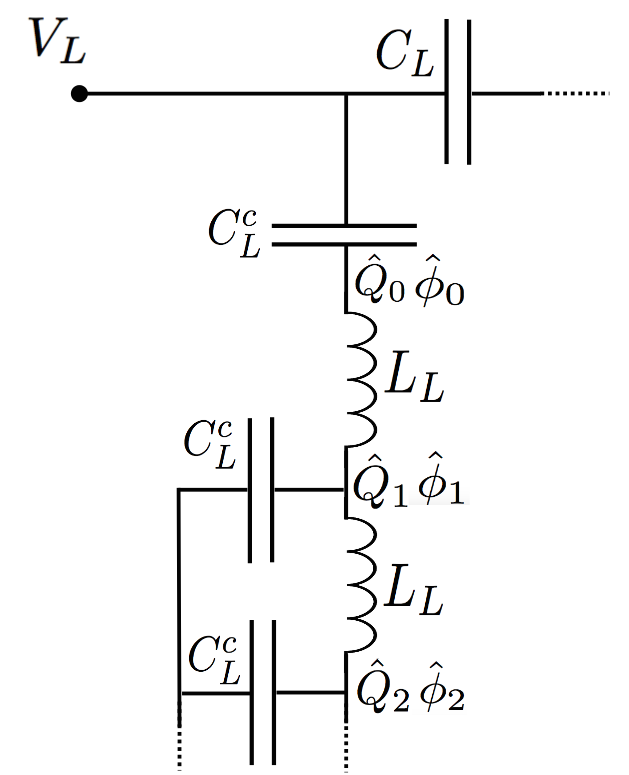}  
\caption{(color online) Left impedance in the case of Fig.~1b. The Left impedance is here a collection of capacitors and inductors. The correspondence between impedance and transmission line just fixes the ratio between inductances and capacitances. We choose the capacitances to be equal to $C_L^c$ and adjust the inductances such that the ratio $\sqrt{L_L/C_L^c}$ gives the resistance at $\omega=0$.}
%\label{fig_6}
\end{figure}
We consider the case pictured in Fig.~8  (coupling between the Left impedance and the voltage fluctuations on the left capacitor), where a resistance is modeled by an infinitely long transmission line. The Hamiltonian of the environment reads \cite{charging_2}
\begin{align}
H_{ph}=\sum_{n=0}^{\infty} \left[ \frac{\hat{Q}_{n}^2}{2C^c_L}+\left(\frac{\hbar}{e}\right)^2 \frac{\left(\hat{\phi}_{n+1}-\hat{\phi}_{n}\right)^2}{2L_L}\right],
\end{align}
where $n$ denotes the site index of the transmission line. Following Ref. \onlinecite{Cedraschi_Buttiker} we can diagonalize the Hamiltonian by putting
\begin{align}
\hat{Q}_n=&\sqrt{2}\int_0^{1} dx \cos \left(\pi n x+\frac{\pi x}{2} \right)\hat{Q}(x)  \notag \\
\hat{\phi}_n=&\sqrt{2}\int_0^{1} dx \cos \left(\pi n x+\frac{\pi x}{2} \right)\hat{\phi}(x).
\end{align}
We can verify that the couple of operators $\hat{Q}(x)$ and $\hat{\phi}(x)$ satisfy the canonical commutation relations. Introducing then creation and annihilation operators $\hat{a}^{\dagger}_x$ and $\hat{a}_x$
\begin{align}
\hat{\phi}(x)=&  \frac{e}{i\sqrt{2 C^c_L \hbar \omega_x} } \left( \hat{a}_x-\hat{a}^{\dagger}_x\right),            \notag \\
\hat{Q}(x)=&  \sqrt{\frac{C^c_L\hbar \omega_x}{2} } \left( \hat{a}_x+\hat{a}^{\dagger}_x\right),                                    
\end{align}
with $\omega_x=2 \sin( \pi x/2)\omega_c$ and $\omega_c=1/\sqrt{L_L C^c_L}$ we get
\begin{align}
H_{ph}=\int_0^1 dx~& \hbar \omega_x \left(a_x^{\dagger}a_x+\frac{1}{2}\right).
\end{align}
%&-\int dx \lambda_x \left(a_x^{\dagger}+a_x\right)\sum_k c_k^{\dagger}c_k,
The voltage fluctuations on the junction capacitor are coupled to the environmental degrees of freedom and we have $e \delta V_L=e\hat{Q}_0/C^c_L=\int dx \lambda_x \left(a_x^{\dagger}+a_x\right)$, where $\lambda_x=e\sqrt{\hbar \omega_x /  C^c_L}\cos (\pi x/2)$. This coupling term is absorbed by the unitary transformation introduced in the main text, and the tunneling element from the left lead to the dot is dressed with the phase factor $\phi_L$ (in the main text) which corresponds to $-\hat{\phi}_0$ using the notations of the Appendix. We have for the time dependent correlation function of the variable $\hat{\phi}_0$

\begin{align}
K_L (t)=\int_0^1 dx \frac{e^2 \cos^2 (\pi x/2)}{\hbar \omega_x C^c_L}\left(e^{-i\omega_x t}-1 \right).
\end{align}
After a change of integration variable from $x$ to $\omega_x$, we finally reach
\begin{align}
K_L (t)=\int_0^{2 \omega_c} 2 \frac{d \omega_x}{\omega_x}  \frac{\sqrt{\frac{L_L}{C^c_L}}\sqrt{1-\left(\frac{\omega_x}{2\omega_c}\right)^2}}{\frac{h}{e^2}}\left(e^{-i\omega_x t}-1 \right).
\end{align}
The behaviour of this function is dominated by the low-frequency modes $\omega_x /\omega_c \ll 1$, so that we can consider equations (3) to be valid with $Z_L=\sqrt{L_L/C^c_L}$.

%\begin{figure}[t!]
%\center
%\includegraphics[scale=0.18]{transmission_line.pdf}  %\includegraphics[scale=0.2]{resonator.pdf}  
%\caption{(color online) Left: Transmission line. Right: Resonator. }
%\label{fig_6}
%\end{figure}
%The case of the coupling to a single-mode resonator corresponds to the right part of Fig.~8.

\end{document}